\documentclass[aps]{revtex4}

\usepackage{epsfig}

\newcommand{\be}{\begin{equation}}
\newcommand{\ee}{\end{equation}}
\newcommand{\bc}{\begin{center}}
\newcommand{\ec}{\end{center}}
\newcommand{\bea}{\begin{eqnarray}}
\newcommand{\eea}{\end{eqnarray}}
\newcommand{\ba}{\begin{array}}
\newcommand{\ea}{\end{array}}
\newcommand{\nn}{\nonumber}

\begin{document}

\title{Analytical obtention of eigen-energies for lens-shaped quantum dot with finite barriers}
\author{Arezky H. Rodr\'{\i}guez$^{1}$, Hanz Y. Ram\'{\i}rez$^{2,3}$}
\affiliation{$^{1}$ Universidad Aut\'onoma de la Ciudad de M\'exico (UACM), Calzada Ermita
Iztapalapa s/n Col. Lomas de Zaragoza, C.P. 09620, Iztapalapa, M\'exico D.F., M\'exico. \\
$^{2}$ Physics Department, Universidad de Los Andes, Carrera 1. No. 18A-10, Bogot\'a, Colombia. \\
$^{3}$Electrophysics Department, National Chiao Tung University, 1001 Ta Hsueh Road, Hsinchu,
Taiwan, R.O.C.}

\date{\today}

\begin{abstract}
The bound states of a particle in a lens-shaped quantum dot with finite confinement potential are
obtained in the envelope function approximation. The quantum dot has circular base with radius $a$
and maximum cap height $b$, and the effective mass of the particle is considered different inside
and outside the dot. A 2D Fourier expansion is used in a semi-sphere domain with infinite walls
which contains the geometry of the original potential. Electron energies for different values of
lens deformation $b/a$, lens radius $a$ and barrier height $V_o$ are calculated. In the very high
confinement potential limit, the results for the infinite barrier case are recovered.
\end{abstract}


\maketitle

\section{Introduction}
\label{intro}

The carrier confinement within small regions such as quantum wells \cite{trallero86}, quantum wires
\cite{trallerito} and quantum dots \cite{hawrylak-book,Grundmann-book} are of a great importance
when in describing transport phenomena, electrical and optical properties of these ``man-made''
systems. Different geometries have been considered (pyramids
\cite{cusack96,cusack97,grundmann95,pryor98,grundmann99}, quantum disks \cite{emp99}, spherical
quantum dots \cite{marin98,emp97,vozumi99}, quantum lenses
\cite{forchel96,hapke99,zhu98,zou99,loualiche2005-1,loualiche2005-2} and even an arbitrary geometry
\cite{patata99}). Due to complex realistic geometries and boundary conditions to include the
effects of the surrounding media, it is not possible in general to find analytical solutions using
common standard procedures. As a first approximation, impenetrable barriers are often considered
since it simplifies the mathematical problem. Nevertheless, the finite value of the potential
barrier could be a fundamental parameter when considering different external potentials or when
including others physical effect, such as the presence of a hydrostatic pressure in a quantum dot
\cite{duque-jpcm}.

When including the finite barrier, different approaches have been used. Bound states in rectangular
cross-section quantum wires as products of eigenstates of 1D problems with a finite barrier in each
direction were found in ref. \cite{califano}. The energy levels are then corrected by the
first-order perturbation-theory. It was shown that the method is suitable for rectangles with
sufficiently large linear dimensions. The same idea was previously applied in \cite{goff} to
calculate the electronic states in cylindrical quantum dots of semiconductors. A 2D Fourier
expansion has been used in \cite{gershoni} to find the electronic states in InGaAs/InP quantum
well-wires structures and in self-assembled InAs pyramidal quantum dots \cite{califano2000}.

Likewise, previous theoretical studies in self-assembled quantum dots with lens shape considered
infinite wall potential \cite{jpacm,jap,PRB-DR}. The aim of the present work is to develop a model
which allows the analytical calculations of the electronic levels in self-assembled quantum dots
with lens shape including a finite barrier potential. The obtained results are compared with those
from considering infinite barrier model and analysis is done establishing the cases where the later
represents a good approximation. The solution of the problem is also found using numerical
calculations for comparing with the analytical results. Finally, some conclusion are outlined.

\section{Model for finite potential}

\begin{figure*}
\vspace*{1cm}
\centerline{\psfig{file=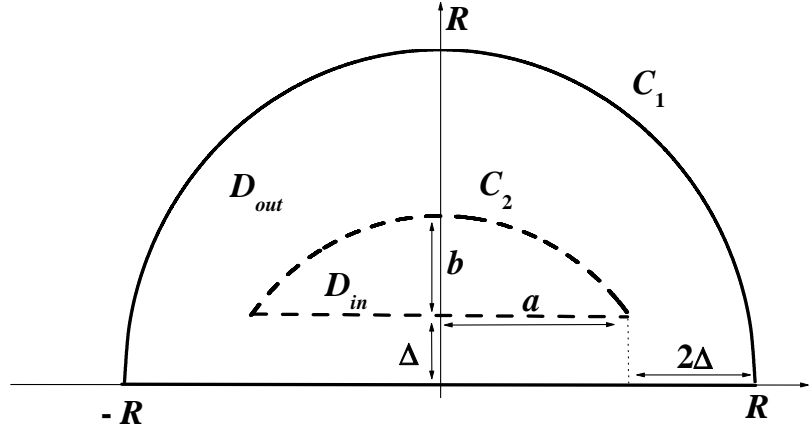,width=0.65\textwidth}}
\vspace*{-4cm}
\caption{Transversal section of a 3D finite lens-well $D_{in}$ with barrier height $V_o$ and contour $C_2$
inside an infinite semispherical-well with contour $C_1$. Contour $C_1$ and $C_2$ are
se\-pa\-ra\-ted a distance equal or greater than $\Delta$ along the perpendicular axis.}
\label{fig:FinBarr}
\end{figure*}

The eigenvalue problem of the Schr\"odinger equation in a 3D lens shape with infinite barriers in
the effective mass approximation has been solved elsewhere \cite{jpacm}. In our case, the problem
for a finite barrier will be modeled including a lens shape well potential with height $V_o$ in a
hard-walls semi-spherical region as shown in Fig. \ref{fig:FinBarr}. The semi-spherical region is
divided in two regions, $D_{out}$ with potential $V_o$ and region $D_{in}$ where it is zero. We
will consider a different value of effective mass for the particle in each region. The solution of
the problem given by the lens with finite barrier in an infinite surrounding medium can be obtained
by minimizing the effect of the external boundary $C_1$ over the wavefunction of the corresponding
energy level under study. This can be achieved by taking a high enough value of the distance
$\Delta$. The equation for the whole region $D$ is given by:
\be
\label{eq1}
- \frac{\hbar^2}{2} \nabla \left( \frac{1}{m^*(r,\theta,\phi)} \nabla \Psi \right) +
V(r,\theta,\phi) \, \Psi = E \, \Psi, \hspace*{1cm} (r,\theta,\phi) \in D = D_{in} + D_{out} \nn
\ee
where
\be
V(r,\theta,\phi) = \left\{
\begin{tabular}{ll}
0 & ;  $(r,\theta,\phi) \in D_{in}$ \\
$V_o$ & ;  $(r,\theta,\phi) \in D_{out}$
\end{tabular}
\right.
\ee
and
\be
m^*(r,\theta,\phi) = \left\{
\begin{tabular}{ll}
$m^*_{in}$ & ; $(r,\theta,\phi) \in D_{in}$ \\
$m^*_{out}$ & ; $(r,\theta,\phi) \in D_{out}$
\end{tabular}
\right.
\ee

The analytical solution of equation (\ref{eq1}) is sought in the form of an expansion
\be
\label{serie}
\Psi = \sum_i C_i \; \Psi^{(o)}_i
\ee
where the set of functions $\{\Psi^{(o)}_i\}$ is a complete set of functions in the 3D domain $D$
given by the semi-sphere. Its explicit representation can be found in \cite{jpacm}, where a
diagonalization procedure was implemented to obtain the electron states. With such conditions, the
functions $\Psi$ satisfy the boundary condition of infinite barrier in the contour $C_1$ because
the set of functions $\{\Psi^{(o)}_i\}$ does. On the other hand, Eq. (\ref{eq1}) and the
corresponding solution given by Eq. (\ref{serie}) are given in the whole domain $D$. It guarantees
that the matching conditions at the contour $C_2$ are also satisfied, but only at those points
where the derivative of the wavefunction is well-defined. This does not occur at the corner and,
generally speaking, the problem is then not well-defined. Then, the obtained eigenvalues constitute
only an estimation of the real problem but this solution constitute a better estimation for the
eigenvalues when the finite barrier is included. This treatment has been applied in \cite{goff} for
a cylindrical domain and in \cite{bruno} for a rectangle, but not explicit analysis was done in the
fulfillment of the matching conditions between the internal and the external domain.

Equation (\ref{eq1}) can be rewritten as:
\be
\label{eq4}
- \nabla^2 \Psi - \sigma(r,\theta,\phi) \nabla \left( \frac{1}{\sigma(r,\theta,\phi)} \right)
\nabla \Psi + \mathcal{V}(r,\theta,\phi) \, \sigma(r,\theta,\phi) \, \Psi = \lambda \,
\sigma(r,\theta,\phi) \, \Psi
\ee
where $\sigma(r,\theta,\phi) = 1$ and $\mathcal{V}(r,\theta,\phi) = 0$ in the internal region
$D_{in}$ and $\sigma(r,\theta,\phi) = \sigma = m^*_{out} / m^*_{in}$ and
$\mathcal{V}(r,\theta,\phi) = \mathcal{V}_o = V_o/E_o$ in the external domain $D_{out}$. The
eigenvalue is given now by $\lambda=E/E_o$ where $E_o = \hbar^2 / 2 m^*_{in} R^2$ is the unit of
energy.

\begin{figure*}
\vspace{3cm}
\centerline{\psfig{file=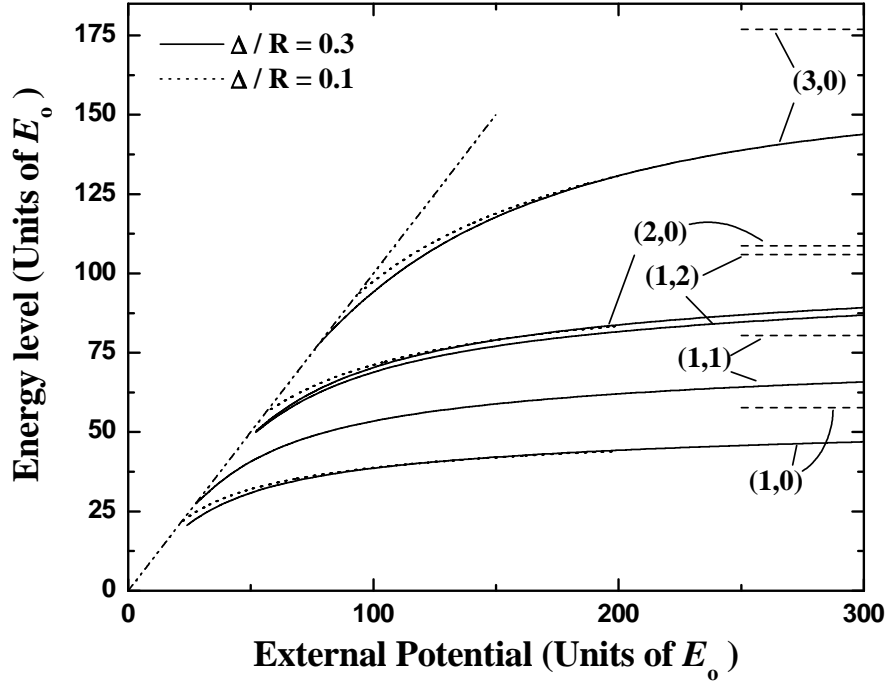,width=0.80\textwidth}}
\vspace*{-3.5cm}
\caption{First five energy levels for a lens-shaped quantum dot with $b/a$=0.51 as
a function of the external potential barrier $\mathcal{V}_o$ in dimensionless quantities. The
calculation is done taking $\Delta/R$=0.3 (solid lines) and $\Delta/R$=0.1 (dotted lines). In both
cases it is used $\sigma = 3.5$. Dashed lines represent the infinite barrier case \cite{jpacm}
while dash-dot-dot is the line where the energy value is equal to the potential value.}
\label{fig:En_adim_Vo}
\end{figure*}

From Eq. (\ref{eq4}) and Eq. (\ref{serie}) it is obtained the matrix representation of the problem
\be
\label{repmat}
\sum_i C_i \left\{ \lambda_i^{(o)} \delta_{ij} - \mathcal{C}(i,j) + \mathcal{A}(i,j) - \lambda \;
\mathcal{B}(i,j) \right\} = 0
\ee
where $\lambda_i^{(o)}$ are the corresponding eigenvalues of the set of functions
$\{\Psi^{(o)}_i\}$. The matrix
\be
\mathcal{C}(i,j) = \left< \Psi_j^{(o)} \left| \sigma \nabla \left( \frac{1}{\sigma} \right) \right.
\nabla \Psi_i^{(o)} \right>_D
\ee
is equal to zero because of the finite discontinuity of $\sigma(r,\theta,\phi)$, and
\bea
\mathcal{A}(i,j) & = & \left< \Psi_j^{(o)} \left| \mathcal{V}(r,\theta,\phi) \,
\sigma(r,\theta,\phi) \right| \Psi_i^{(o)} \right>_D = \mathcal{V}_o \, \sigma \, \left[
\delta_{i,j} - \left< \Psi_j^{(o)} | \Psi_i^{(o)} \right>_{D_{in}} \right]
\\
\mathcal{B}(i,j) & = & \left< \Psi_j^{(o)} | \sigma(r,\theta,\phi) | \Psi_i^{(o)} \right>_D =
\left< \Psi_j^{(o)} | \Psi_i^{(o)} \right>_{D_{in}} + \; \sigma \, \left[ \; \delta_{i,j} \; - \;
\left< \Psi_j^{(o)} | \Psi_i^{(o)} \right>_{D_{in}} \right]
\eea
where $<>_{D_{in}}$  means an integration over the internal domain $D_{in}$.

According to the axial symmetry, the Hilbert space of the problem given by Eq. (\ref{repmat}) is
separated in different subspaces, each one characterized by a quantum number $m$. The first five
eigenvalues $\lambda$ for $b/a$=0.51 as a function of the external potential $\mathcal{V}_o$ are shown in Fig. \ref{fig:En_adim_Vo}. Each one is labeled by a couple of indexes $(N,m)$ meaning the $N$-th energy level with axial quantum number $m$. It is used $\sigma = m^*_{out} / m^*_{in} = 3.5$ which is the ratio between the values of the effective masses in an InAs/GaAs quantum dot material \cite{duque-jpcm}. It can be seen that, as the
external potential increases, also increase the energy levels approaching asymptotically to the
corresponding values of the infinite potencial case which are shown horizontally in dashed lines
\cite{jpacm}. For a given value of the potential barrier, the energy values for the lower levels
are closer to the corresponding value taken the barrier as infinite than those for higher levels,
as expected. At the same time, as higher the level, higher the percent of the wavefunction located
at region $D_{out}$ and stronger the influence of the artificial boundary $C_1$. This influence is
also stronger for lower values of $\Delta/R$. This effects can be seen in Fig. \ref{fig:En_adim_Vo}
when comparing the solid lines, calculated by using $\Delta/R$=0.3, with the dotted lines,
calculated by using $\Delta/R$=0.1 (only for levels with $m=0$). However, at those values of
the potential barrier where the solution is independent of the parameter $\Delta/R$, the solution
can be taken as independent of the boundary $C_1$ and hence, as a good approximation for the finite
barrier case in an infinite surrounding medium. Dash-dot-dot line represents the points where the
energy value is equal to the potential value.

\begin{figure*}
\centerline{\psfig{file=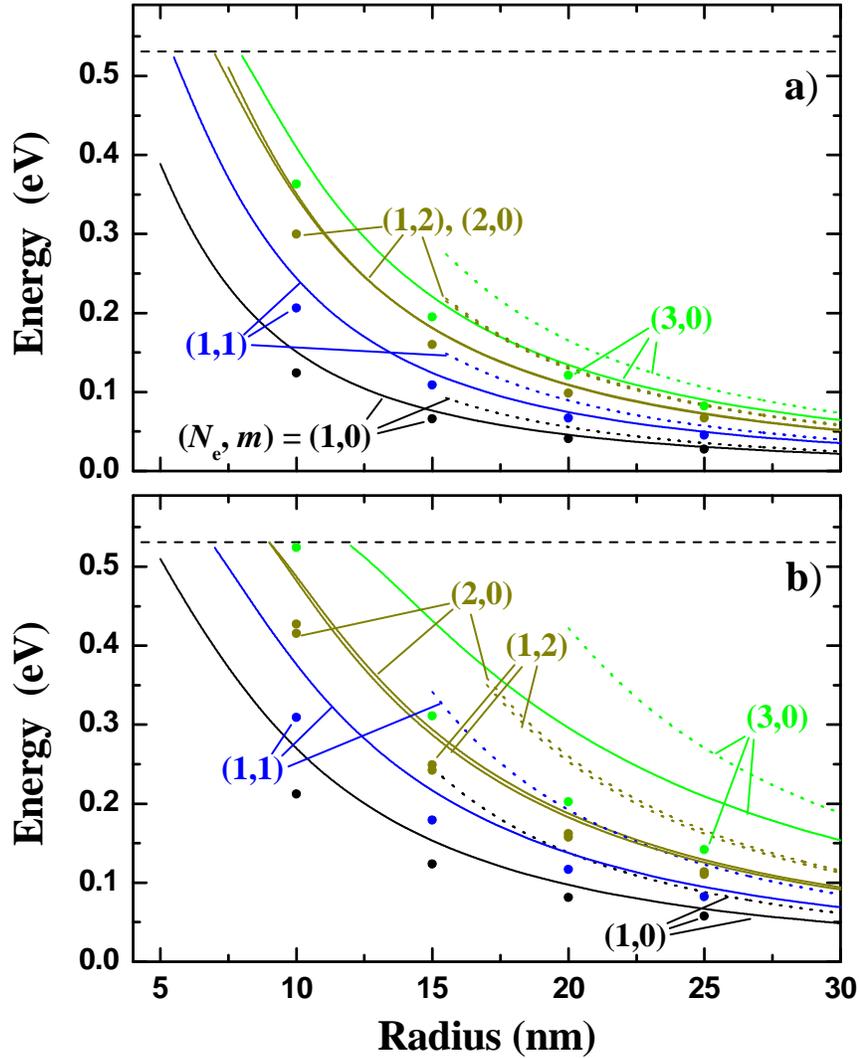,width=0.70\textwidth}} \vspace*{-1cm}
\caption{First five electronic levels for an InAs/GaAs quantum lens as a function of the lens radius.
a) $b/a$=0.91, b) $b/a$=0.51. The calculation is done taking $\Delta/R$=0.3 (solid lines).
Results from the infinite barrier model are also plotted for comparison (dotted lines). As a
reference, the value for the confinement potential used ($V_{c}=0.531$eV) is shown by dashed line
in both panels. Filled dots correspond to numerical calculations.}
\label{fig:Edim_radio}
\end{figure*}

In order to study a particular quantum dot material, two different quantum lens configuration of
InAs/GaAs have been considered. In Fig. \ref{fig:Edim_radio} the first five electronic levels are
shown as a function of the lens radius, where a 500x500 matrix was used in the diagonalization
procedure. The material parameters used here for the calculation are the same as in
\cite{duque-jpcm}.

In general, the values of the energy levels decrease for increasing values of the radio. As shown
in Fig. \ref{fig:Edim_radio} a), for $b/a$=0.91 and according to the levels shown, the infinite
barrier model is a good approximation for radius of the order of 20 nm or higher. Nevertheless, as
seen in Fig. \ref{fig:Edim_radio} b), for lower values of $b/a$ it is necessary to include the
finite barrier effects to get better approximations of the energy levels distribution for all the
values of the radius shown.

As an intend of verifying the obtained results, numerical calculations were carried out solving
directly the BenDaniel-Duke equation of the system, calculating the eigenvalues by using the finite
elements technique through programs for Comsol application, as used in previous works
\cite{Hanz,Hanz2}. The corresponding results are shown by filled dots in Fig. 3, calculated with
the same material and geometrical parameters as those used in the analytical curves. Although
qualitatively the behavior of the analytical and numerical results are consistent, since the
quantitative point of view the numerically obtained values have always lower values than those
represented by the solid and the dotted lines. Furthermore, the tree models coincide for higher
enough values of the dot radius, but its results become different when the radius decreases. The
result obtained is mainly due to the presence of the frontier $C_1$ (at the analytical calculation)
whose effects become important for smaller dots because of the increasing of the energy values and
correspondingly, the wavefunction has higher percent outside the lens domain given by $D_{in}$ in
Fig. \ref{fig:FinBarr}. In the same way, the necessary basic truncation introduce an error which
becomes important for smaller dots and, as found in \cite{jpacm}, the accuracy of the analytical
method requires bigger matrices for larger lens deformation (smaller values of $b/a$), which is in
agrement with the comparison of panels a) and b) from Fig. 3.

\section{Conclusions}

In the present work the results from \cite{jpacm} and \cite{jap} have been generalized to evaluate
the electronic energies in self-assembled quantum dots with lens shape geometry taking into account
the finite barrier height. The results obtained by the present model was compared with the values
obtained when considering the potential barrier as infinite and with a numerical calculation
procedure. It was established the range of values for the potential barrier, lens deformation $b/a$
and lens radius $a$ where all the models produce similar results. It was also argue the reasons for
its different energy values obtained for smaller dots and for stronger lens deformations. The
present model can be applied to study analytically the electronic properties of a self-assembled
quantum dots with lens shape under the presence of external potentials where it could be important
to consider the actual values of the finite barrier.

\section*{Acknowledgements}

AHR thanks many valuable discussions with Dr. C. Trallero-Giner.

\end{document}